\documentclass[1p,12pt]{elsarticle}
\usepackage{amssymb}
\usepackage{graphicx}
\usepackage{geometry}
\usepackage{siunitx}
\DeclareSIUnit\mil{mil}
\DeclareSIUnit\ounce{oz.}
\journal{New England iMAPS for the 2014 Symposium \& Expo}
\begin{document}
\begin{frontmatter}
\title{Design and Fabrication of a Highly Integrated Silicon Detector for the
STAR Experiment at Brookhaven National Laboratory}
\author[mit]{Benjamin~Buck\corref{cor_au}}
\ead{bbuck@mit.edu}
\author[lbl]{Eric~Anderssen}
\author[mit]{Jason~Bessuille}
\author[lbl]{Mario~Cepeda}
\author[lbl]{Thomas~Johnson}
\author[mit]{James~Kelsey}
\author[mit]{Gerrit~van~Nieuwenhuizen}
\author[iu]{Gerard~Visser}
\address[iu]{Indiana University, Bloomington, IN, USA}
\address[lbl]{Lawrence Berkeley National Laboratory, Berkeley, CA, USA}
\address[mit]{Massachusetts Institute of Technology, Cambridge, MA, USA}
\cortext[cor_au]{Corresponding author}
\begin{abstract}
We present the design of a detector used as a particle tracking device in the
STAR experiment at the RHIC collider of Brookhaven National Laboratories. The
``stave,'' 24 of which make up the completed detector, is a highly mechanically
integrated design comprised of 6 custom silicon sensors mounted on a Kapton
substrate. 4608 wire bonds connect these sensors to 36 analog front-end chips
which are mounted on the same substrate. Power and signal connectivity from the
hybrid to the front-end chips is provided by wire bonds. The entire circuit is
mounted on a carbon fiber base co-cured to the Kapton substrate. We present the
unique design challenges for this detector and some novel techniques for
overcoming them.
\end{abstract}
\end{frontmatter}
\section{Introduction}
The Intermediate Silicon Tracker (IST) is a silicon based particle detector
installed at Brookhaven National Laboratory (BNL) in the Relativistic Heavy Ion
Collider (RHIC). The detector is located in the central region of the
Solenoidal Tracker at RHIC (STAR) experiment
and makes up the third tracking layer in a 4-layer vertex detector
upgrade. It is comprised 24 staves, each of which is a flexible PCB wrapped
around a carbon fiber core. Attached to each stave are 6 silicon sensors which
can detect energetic particles which pass through them. These silicon sensors are read
out by 36 analog front-end chips which are connected to an external data acquisition system.

\section{Hybrid design}
\begin{figure}[h]
\begin{center}
\includegraphics[width=5in, keepaspectratio=true, angle=0]{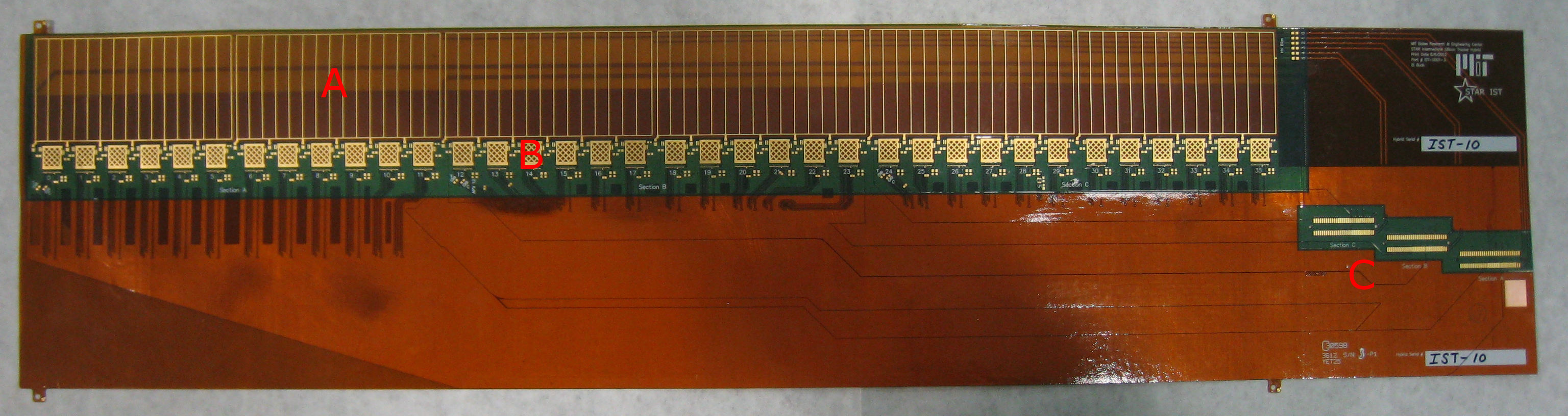}
\caption{Image showing one of the IST hybrids. `A' marks a sensor placement site,
`B' marks an APV placement site, `C' marks a connector placement site.
\label{fig:hybrid}}
\end{center}
\end{figure}

Each stave holds the electrical substrate known as a ``hybrid.''  The hybrid is
the flexible circuit onto which the silicon sensors, analog front-end chips, and passive
components are mounted. The hybrid is permanently bonded to the stave during
the stave manufacturing. The hybrid is broken up into 2 areas, the sensor area
and the connector area. In the sensor
area we desire as little mass as possible to reduce interactions with passing
particles. For this reason, the sensor area, which makes up most of the hybrid,
is made of 2 layers of \SI{0.5}{\ounce} copper on a \SI{1}{\mil} Kapton substrate. The connector
area is made of the same 2 layers of \SI{0.5}{\ounce} copper but with an additional
\SI{1}{\ounce} copper layer and \SI{1}{\mil} Kapton layer. Over both areas is a \SI{1}{\mil} Kapton
coverlay attached with \SI{1}{\mil} of adhesive. This coverlay has no cutouts on the
bottom layer and acts as insulation between the copper and the structural
material in the stave. On the top layer large areas of this cover lay are cut out around
any area with bonding pads or component pads. Conventional solder mask is
applied to this area as shown in figure~\ref{fig:hybrid}. The hybrids are finished with ENEPIG for compatibility
with both wire bonding and conventional soldering. In order to keep the copper
weight down, all vias are selectively plated through.  This selective plating
can be seen in figure~\ref{fig:passive}.

Each hybrid provides the electrical connection from the APV chips to the
connector site on one end of the stave. Electrically the hybrid is broken up
into three separate circuits which share a common ground. Each circuit
serves 12 APV chips and 2 sensors. The APV chips require power, analog output,
and control signals to be routed to each chip. The sensors require a bias line
which must withstand up to \SI{200}{\volt}.

In the sensor area we require as little mass as possible. Because of this, the sensor area has
only two routing layers, so careful routing was needed to optimize the power net
distribution and signal distribution. The power nets are distributed as three
large busses per section for \SI{+1.25}{\volt}, \SI{-1.25}{\volt}, and ground. The power supplies
for the staves operate on a remote regulation scheme, so a sense net from each bus is
routed locally away from the APV chips in each section and back to the
connectors. For
the analog output signals, a differential pair from each APV chip is routed to the
connectors. The analog output is a \SI{100}{\ohm} differential current mode signal. Source
termination is provided at the board which plugs into the stave. In addition,
the APV chips use I2C for slow controls to set various parameters of the chip.
These nets are also routed as a bus to each of the chips in a section. Finally,
two differential pairs provide a clock and a trigger to each APV chip. These
are routed as a bus to each of the chips in a section. 

\begin{figure}[ht]
\begin{center}
\includegraphics[height=1.7in, keepaspectratio=true, angle=0]{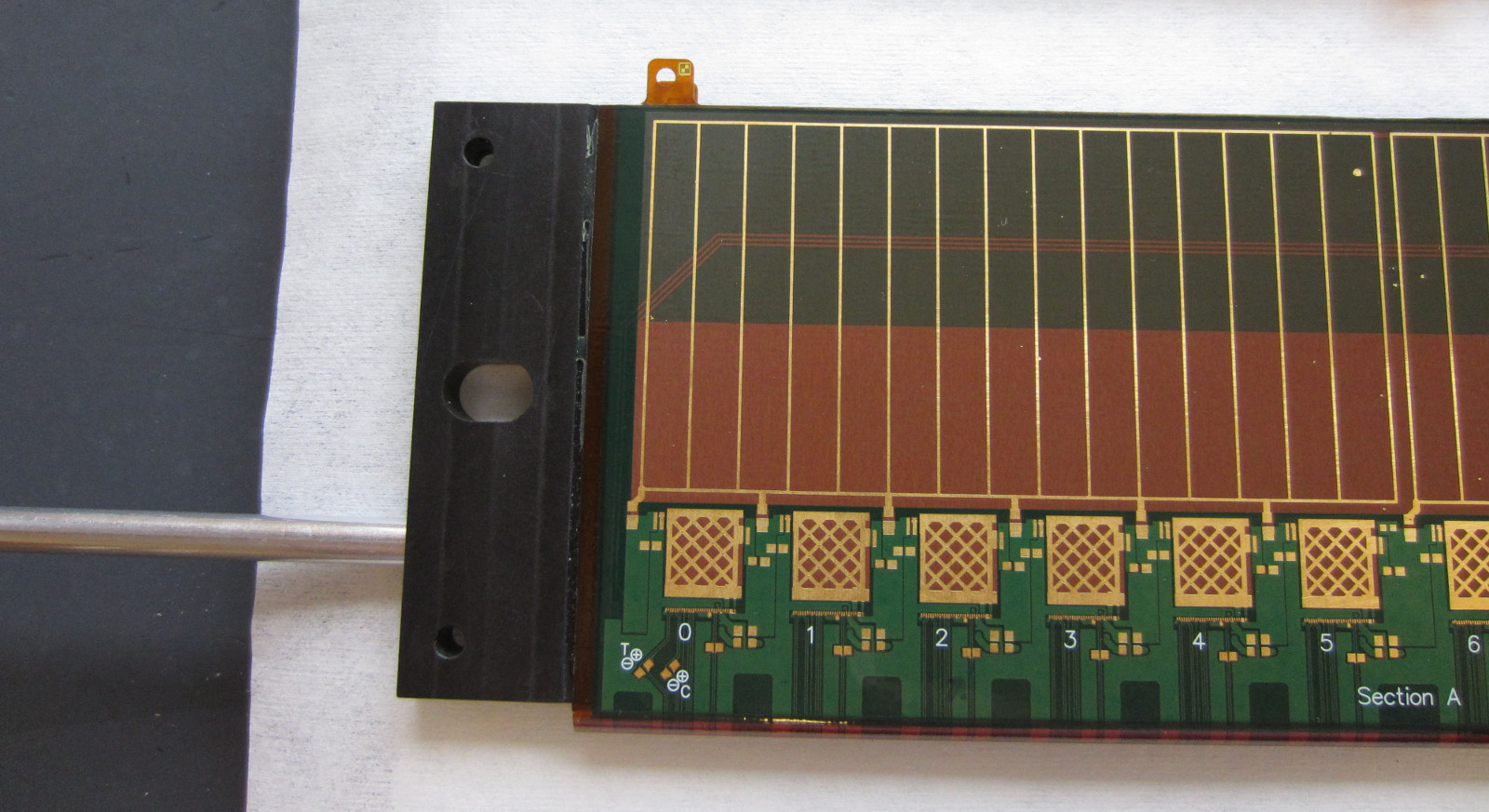}
\includegraphics[height=1.7in, keepaspectratio=true, angle=0]{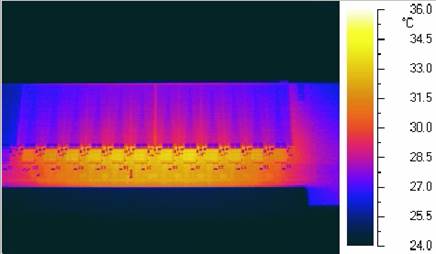}
\caption{Left shows the cooling tube running directly beneath the APV chips which generate nearly
all of the heat on the stave.  Right shows a thermal image of a completed stave
running in still air with no cooling.
\label{fig:cooling_tube}}
\end{center}
\end{figure}

The hybrid has 36 APV placement sites where APV chips are attached and 6 sensor
placement sites where sensors are attached. The APV chips generate nearly all of
the heat in the stave so an aluminum cooling tube is embedded in the stave
directly beneath the APV chips as shown in figure~\ref{fig:cooling_tube}. To reduce the thermal resistance, 14 vias are placed on the APV
placement sites to conduct heat through the Kapton core. The 6 sensors do not
produce any appreciable thermal load.

\section{Stave design}
Each hybrid was assembled into a stave at the Lawrence Berkeley National Laboratory (LBNL) Composites Shop in Berkeley,
California. Two carbon fiber cover sheets were co-cured to a hybrid and then a
carbon fiber honeycomb was glued to one side. A block of carbon foam with a
channel cut in it for cooling was also glued on. The carbon foam has a lower
radiation length but is homogeneous unlike the carbon fiber honeycomb. This
uniformity makes it more suitable for wire bonding, so the carbon foam is
located underneath any area where wire bonding will be performed.

Each stave nominally generates about \SI{12}{\watt} of heat. Cooling is provided by flowing a synthetic heat transfer fluid, 3M\textsuperscript{TM}
Novec\textsuperscript{TM} 7200 Engineered Fluid,
through the aluminum tube which runs through the carbon
foam. Because the IST is installed in the central region of STAR, we needed a
cooling fluid which would not damage the other detectors surrounding it in the
event of a leak. Novec has a very low vapor pressure, leaves no residue when it
evaporates and has a very high resistivity, making it a good candidate. Novec was chosen because it has a
lower ozone-depletion potential than other engineered fluids with similar
properties.  Other detectors using a similar design have observed galvanic corrosion
between the carbon foam and the aluminum tube \cite{ref:novec}. In order to prevent this, the
carbon foam channel was coated with epoxy before the tube was laid in to provide a
barrier. The target running temperature is \SI{25}{\degreeCelsius} which is above the
dew point in the experimental hall; this will prevent moisture from condensing
on the detector which could facilitate galvanic corrosion. In addition, all
parts of the stave and cooling system are electrically grounded.

Two end pieces called ``closeouts'' were then attached on either end of the stave. One
is made of carbon filled PEEK and the other of aluminum. The closeouts have
holes and slots to provide a way to mount the stave to the finished detector. The
aluminum closeout is electrically bonded to the carbon face sheets, the carbon
fiber honeycomb, and the carbon foam. This provides a means to ground the
materials in the stave since they are electrically isolated from the hybrid.

\begin{figure}[h]
\begin{center}
\includegraphics[width=5.5in, keepaspectratio=true, angle=0]{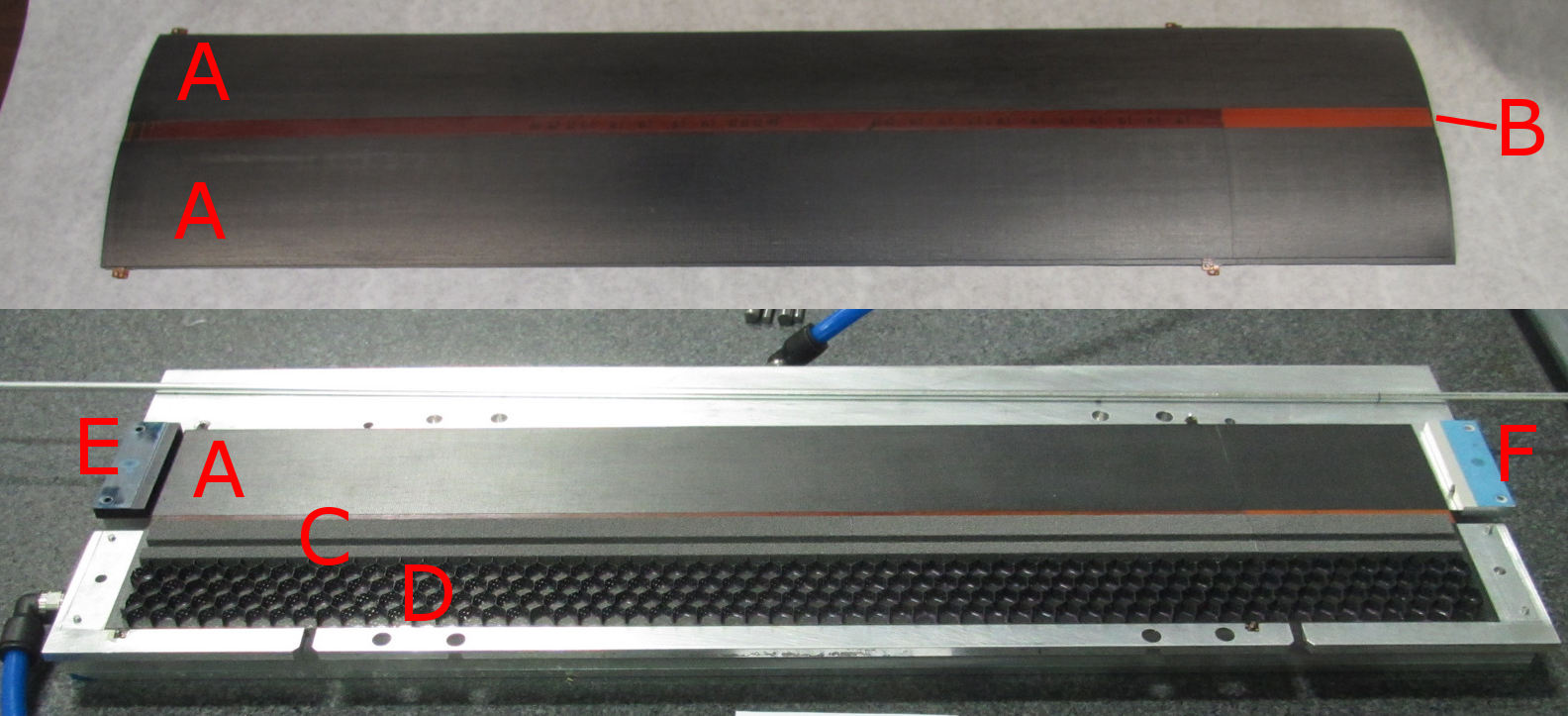}
\caption{Top image shows the hybrid, `B', with the two carbon fiber face sheets, `A', attached.
The bottom image shows the same hybrid now with the carbon fiber honeycomb, `D', and 
carbon foam, `C', attached. The PEEK closeout, `E', and aluminum closeout, `F', can also be seen.
\label{fig:stave_assmb}}
\end{center}
\end{figure}

Once everything was bonded to one of the face sheets, glue was applied to
the other side and it was folded over \SI{180}{\degree} to complete the stave.
Figure~\ref{fig:stave_assmb} shows the stave right before this step. Fully
cured staves were inspected and then sent to MIT for further assembly.

\section{Sensor design}
\begin{figure}[h]
\begin{center}
\includegraphics[width=4in, keepaspectratio=true, angle=0]{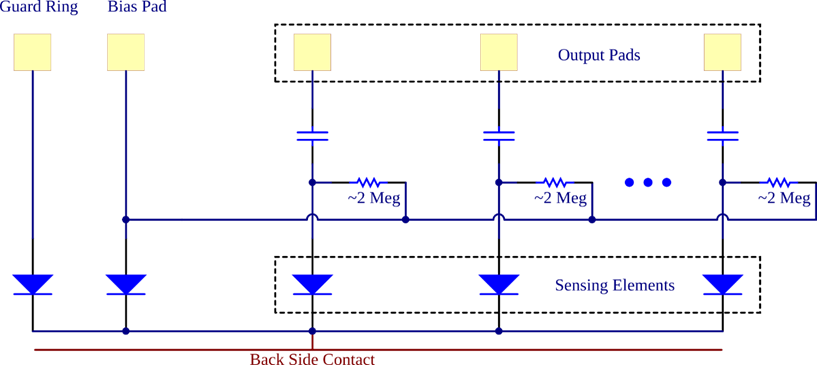}
\caption{A simplified schematic of the silicon sensor.  Charge pairs produced in
the sensing elements are detected on the output pads.
\label{fig:sensor_sch}}
\end{center}
\end{figure}

The IST sensors were designed by MIT and
manufactured by Hamamatsu. The sensors are \SI{300}{\micro\meter} silicon with 2 metallization
layers. The backside has a sputtered aluminum coating to provide the bias voltage.
This voltage reverse biases the PN junctions in the sensor which form the individual
sensing elements (see figure~\ref{fig:sensor_sch}. Particles passing through the sensor will interact
with the silicon lattice and produce charge pairs which will drift apart. A capacitor
above each sensing element AC couples the signal and routes it to a bonding pad
located on one edge of the chip. This is a proven, standard technology for
particle detection.

\begin{figure}[ht]
\begin{center}
\includegraphics[width=3in, keepaspectratio=true, angle=0]{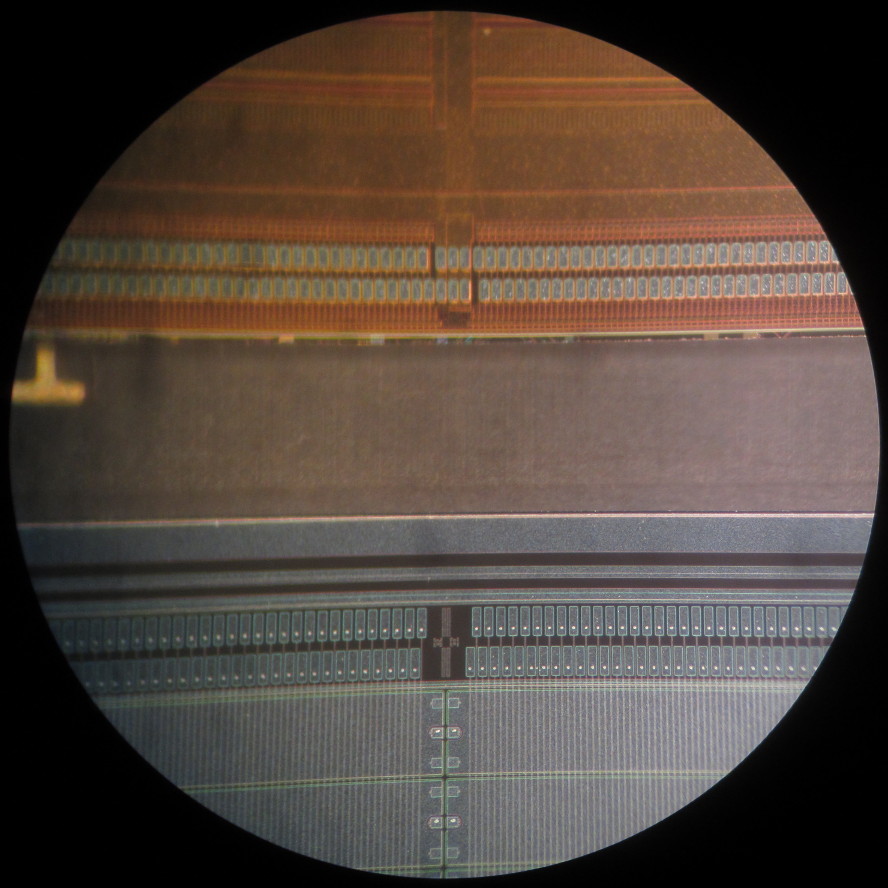}
\caption{APV bonding pads on top and silicon sensor bonding pads on bottom.
\label{fig:bonding_pads}}
\end{center}
\end{figure}

Each sensor has 6 groups of sensing elements with 128 sensing elements in each
group. A bond pad for each one of the sensing elements is grouped at one edge
of the sensor and is arranged in a pattern which mirrors the bonding pattern for
the APV chip (see figure~\ref{fig:bonding_pads}). This allows for much simpler bonding between the APV and the
sensor, especially useful because the bonding pitch is so small. The sensing
elements are arranged in a grid at a pitch of \SI{6275}{\micro\meter}~$\times$~\SI{596}{\micro\meter} in 64 rows and 12
columns. The bonding pads are arranged in 2 rows of 64 pads. The pads are
\SI{58}{\micro\meter}~$\times$~\SI{136}{\micro\meter} with \SI{30}{\micro\meter} between each pad and \SI{80}{\micro\meter} between the 2 rows.

\section{APV chip}
The APV25-s1 chip is the readout and pre-amplifier ASIC for the sensors. It has
128 channels each with a charge sensitive pre-amplifier, shaper, and \SI{4}{\micro\second} long
pipeline. Events are read into the pipeline at \SI{40}{\mega\hertz}. Events in the pipeline are selected by
triggers and marked for readout. A single differential pair per chip reads out each of
the 128 channels in series for a selected event.

\begin{figure}[ht]
\begin{center}
\includegraphics[height=2in, keepaspectratio=true, angle=0]{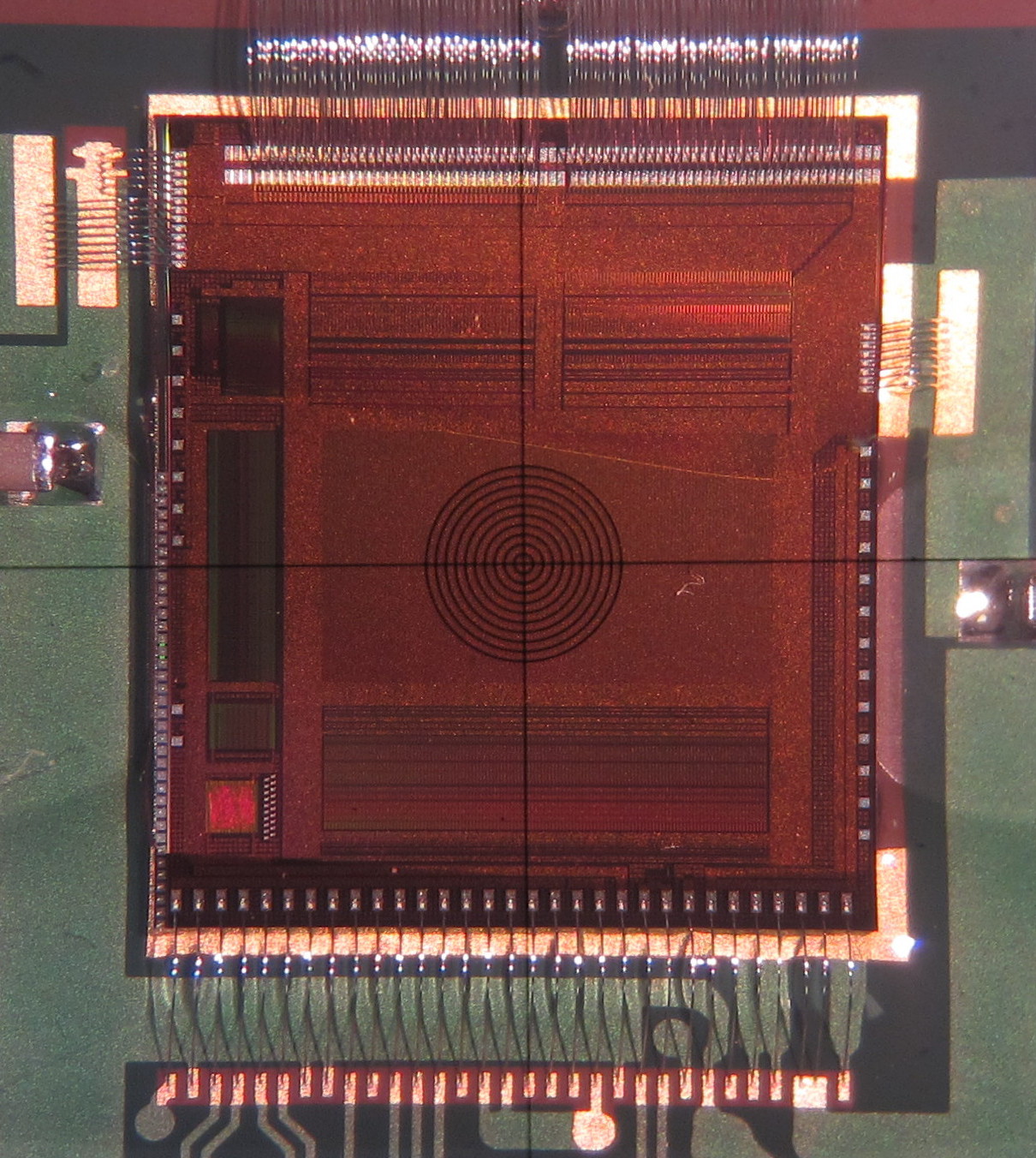}
\caption{APV chip showing input bonding pads on the top, power bonding pads on
the left and right, and control bonding pads on the bottom.
\label{fig:apv}}
\end{center}
\end{figure}
This ASIC was designed by the Imperial College in London for the Compact Muon
Solenoid (CMS) running at the Large Hadron Collider (LHC)~\cite{ref:apv_nim}. It was designed
for the high radiation environment present in the middle of a physics experiment
and used an IBM \SI{0.25}{\micro\meter} radiation hard process. Detailed information about the
APV chip can be found in~\cite{ref:apv} and~\cite{ref:apv_nim}.  Figure~\ref{fig:apv} shows the APV 
chip bonded to a stave.

\section{Passive attach}
In addition to the sensors and APVs, a few passive components are needed on the stave.
Each stave has 195 additional components, mostly 0402 sized capacitors for
bypassing the APV chips. In addition, there is a small temperature sensor,
termination resistors for the clock and trigger lines, protection resistors for
the sensor bias lines, and a connector for each section.

The connectors used were the Samtec TEM and SEM series. They provide a low
mating height and high pin density which is ideal for our application. The
pins are extremely fragile, however, so extreme care had to be taken during
testing and assembly to reduce the number of mating cycles and prevent any
connectors from being damaged.

\begin{figure}[ht]
\begin{center}
\includegraphics[width=3in, keepaspectratio=true, angle=0]{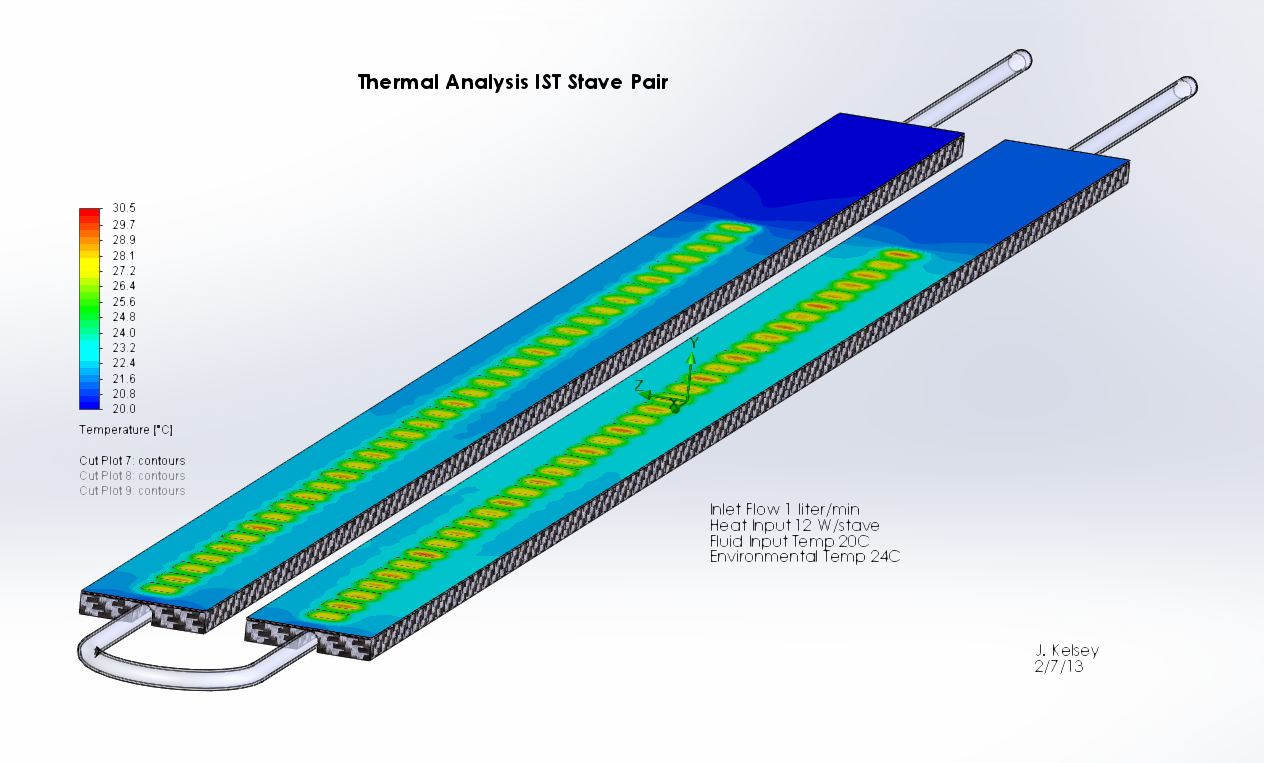}
\caption{CFD study showing two staves which make up a single cooling loop.
\label{fig:cfd}}
\end{center}
\end{figure}

The temperature sensor used was a Texas Instruments TMP102 which shares the same
I2C bus as the APV chips in one of the sections. A computational fluid dynamics (CFD) study was done to
determine the location of the hottest spot on a stave assuming nominal operating
conditions (see figure~\ref{fig:cfd}). The temperature sensor is placed in this location.

\begin{figure}[h]
\begin{center}
\includegraphics[width=3in, keepaspectratio=true, angle=0]{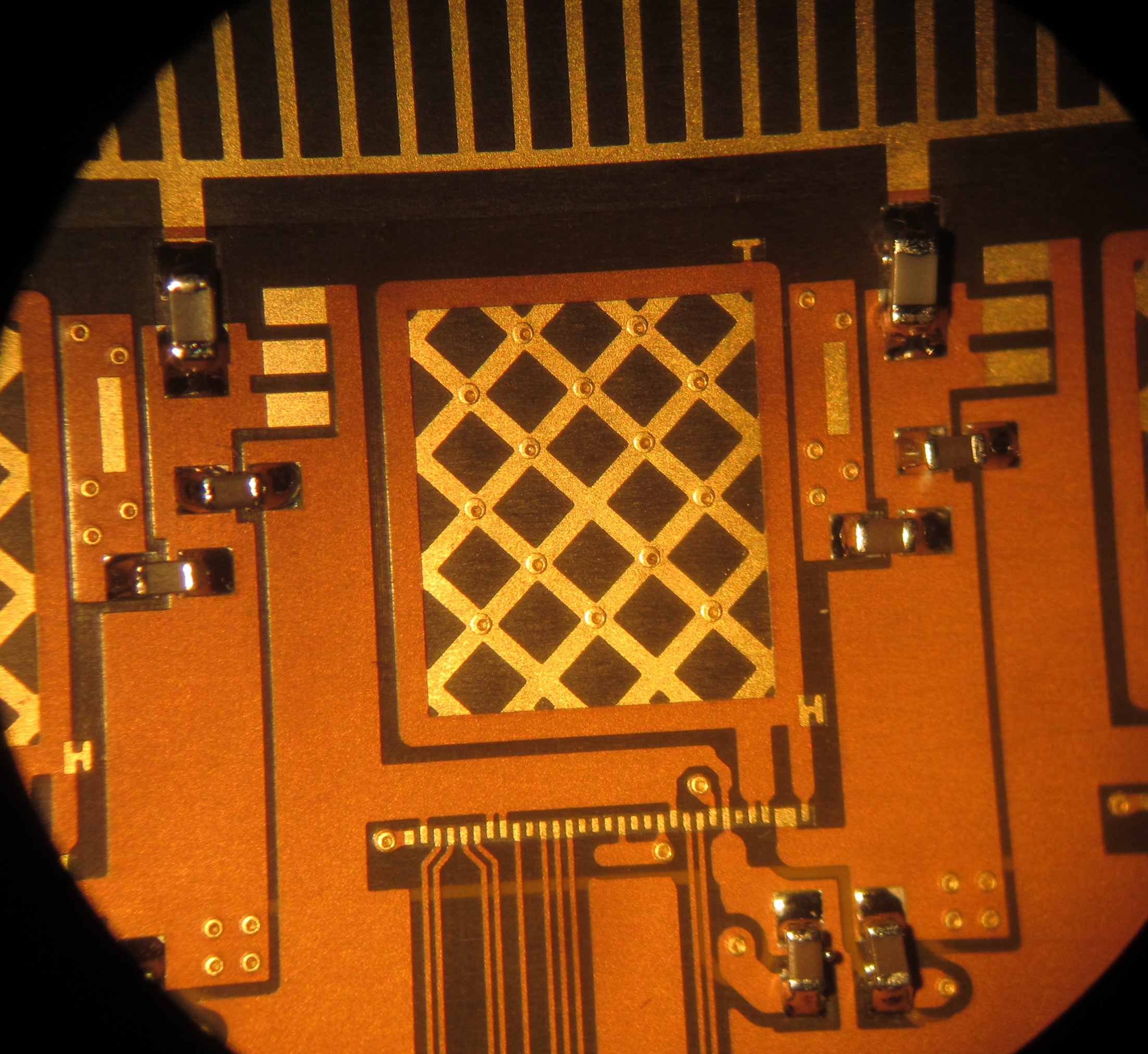}
\caption{Passives attached to stave before APV or silicon sensor attach.  Vias
for thermal conduction under APV chip can also be seen.
\label{fig:passive}}
\end{center}
\end{figure}

The passives were soldered onto the hybrids after stave assembly. Because the
hybrid was already attached to the carbon fiber parts of the stave, it
could not be put into a solder oven and had to be hand-assembled. Proxy
Manufacturing in Methuen, Massachusetts assembled all of the staves. The
assembly process was further complicated because there is no silk screen
legend for most of the components on the board. Because the bypass capacitors
are located so close to the APV chips there was no space for a silk screen
legend (see figure~\ref{fig:passive}). Instead, the pattern of component placement was made the same for all
APV placement sites and detailed documentation was produced to prevent any parts
from being incorrectly placed. A visual inspection at MIT after the
passives were attached to the staves provided an additional quality check for
such an error.

\begin{figure}[h]
\begin{center}
\includegraphics[width=3in, keepaspectratio=true, angle=0]{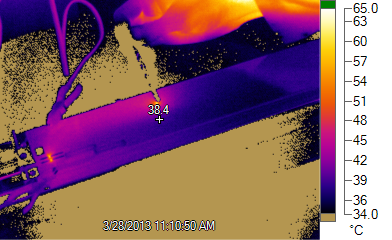}
\caption{Hot area on the left is the connection point to the power supply.  Hot area
in center is location of a short inside the hybrid.
\label{fig:thermal}}
\end{center}
\end{figure}

During production one stave was found to have a short between two power nets on
the hybrid, away from any solder areas. Using a thermal camera we were able to
determine the location of the short (see figure~\ref{fig:thermal}). Until that point in production, the staves were not
electrically tested after receipt from LBNL or after passives were attached.
After this incident, the power nets were electrically tested for shorts. We found no
additional problems during production.

\section{APV attachment}
During production we performed separate assembly steps for the APV chips and the
silicon sensors. Our production volume was limited by the silicon sensors as they were
the single most expensive component to purchase and also had the longest lead
time. Running out of sensors late in production would have delayed installation
of the full detector. To increase the overall yield, the APV chips were
installed on the staves first, wire bonded, and tested to make sure they were
fully functional. Only after all APV chips on a stave were verified to be
working properly were silicon sensors attached.

The APV chips were glued directly to the hybrid using a conductive epoxy, TRA-DUCT
2902. A pneumatic dispenser was used to apply a uniform amount of glue for each
APV chip. The glue is both thermally and electrically conductive and acts as a
means to connect the back side to VSS and as a means to conduct heat away from
the chip.

\begin{figure}[h]
\begin{center}
\includegraphics[width=5in, keepaspectratio=true, angle=0]{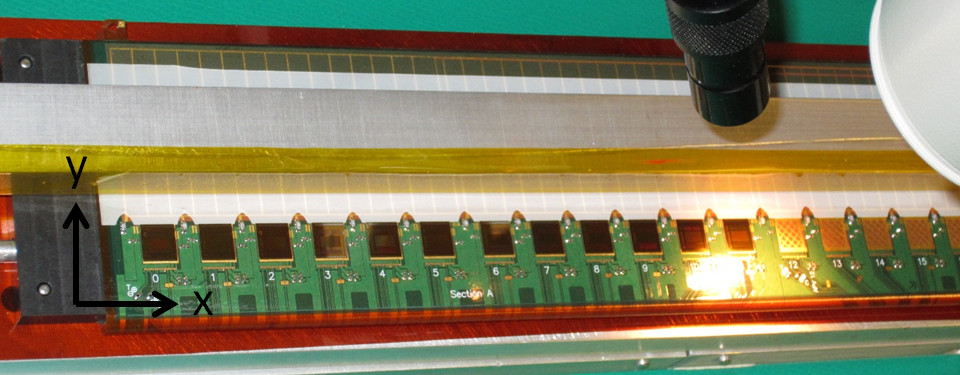}
\caption{APV chips are pressed up against the white guide to provide ``y'' coordinate
alignment.  A technician uses a microscope to align the chips in the ``x'' coordinate.
\label{fig:apv_attach}}
\end{center}
\end{figure}

It is important that the alignment of the APV chips is precise. Since there are
bonding pads on all sides of the chip, both ``x'' and ``y'' alignment are important.
To achieve good alignment we made custom tooling which would allow the
technician to align each chip manually. The tooling provides an edge which all
APV chips are pressed against to set the ``y'' alignment. Then the technician
aligns the bonding pads on the APV chip with copper features on the hybrid to
ensure a precise alignment in ``x''. The tooling used for this is shown in
figure~\ref{fig:apv_attach}. This alignment step was most important
because the mounted APV chips would then have to match exactly the location of
the output pad sites on the silicon sensors. Misalignment would make the bonding
process more difficult and increase the likelihood of shorts between wire bonds.
All of the alignment was done by hand under a microscope by a skilled
technician.

\section{APV bonding and testing}
\begin{figure}[h]
\begin{center}
\includegraphics[width=2.5in, keepaspectratio=true, angle=0]{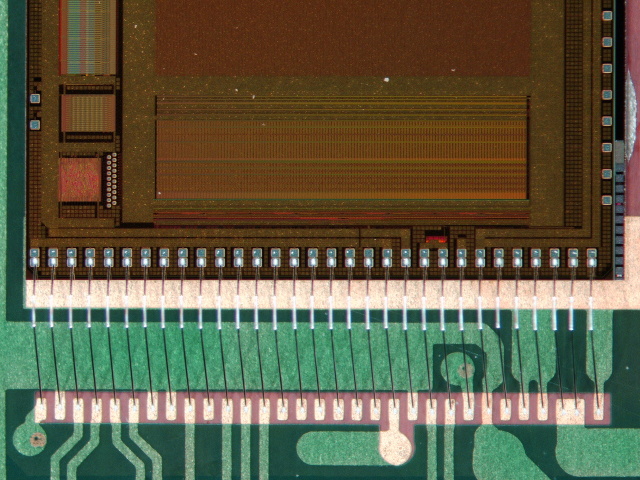}
\includegraphics[width=2.5in, keepaspectratio=true, angle=0]{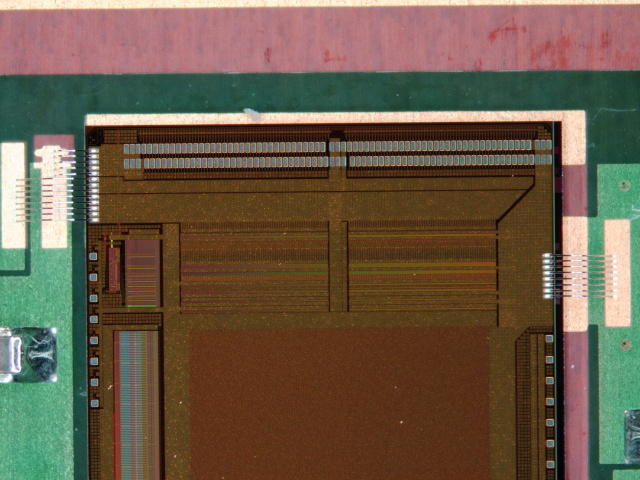}
\caption{Close up of freshly made bonds on the APV.  Control bonds on the left
and power bonds on the right.
\label{fig:apv_bond}}
\end{center}
\end{figure}

Staves with APV chips attached were sent to the Instrumentation Division at
Brookhaven National Laboratory for wire bonding and testing. Wire bonding was
done on a Hesse \& Knipps Bond Jet 815 bonding machine using \SI{1}{\mil} aluminum
wire. A program for the bonding machine was written to wire bond each APV chip
automatically, but inspection was done by a bonding technician during the
bonding process to ensure the operation went smoothly.  Example images from the
inspection are shown in figure~\ref{fig:apv_bond}. During this
operation 31 control nets were bonded from the hybrid to the APV chip and 21
power nets are bonded from the hybrid to the APV chip. There were a total of 1872
wire bonds for this operation. The stave was held down using only the stock
vacuum chuck on the bonding machine. No additional fixture was needed.

Bonded APV chips had to be tested before silicon sensors could be attached. In an
experimental assembly building located next to the STAR detector we set up a
clean room with a testing station for the IST. The testing station consists of
a rack to hold detectors with a cosmic ray trigger inside a light-tight box
(ambient light will create noise on the sensors due to the photoelectric effect.)
The staves were attached to a data acquisition system (DAQ) (see~\cite{ref:daq} for more information)
which provided power and control signals and digitized the data from the APV
chips. To test each stave, it was attached to the DAQ and operated. Noise and pedestal levels
were recorded over many events and then analyzed.

All staves were then sent back to MIT for further work. Staves that had APV
chips on them which did not pass the noise and pedestal level tests had
those chips removed and replaced and then sent back to BNL. The replaced APV
chips were bonded and then re-tested before they were sent back to MIT. Staves
with fully functioning APV chips then had the silicon sensors installed.

\section{Silicon sensor attachment}
The silicon sensor attachment was the final assembly step done at MIT. Before the
process began, all tooling and staves were cleaned with isopropyl alcohol to
ensure that no contaminants were transferred onto the sensors. Staves were placed
in MIT-made tooling which held the stave in place. An adjustable blade at the
top edge of the stave was moved until it marked the top edge of where each sensor
would sit. The blade was fastened in position and then the sensor attach
began.

\begin{figure}[h]
\begin{center}
\includegraphics[width=6in, keepaspectratio=true, angle=0]{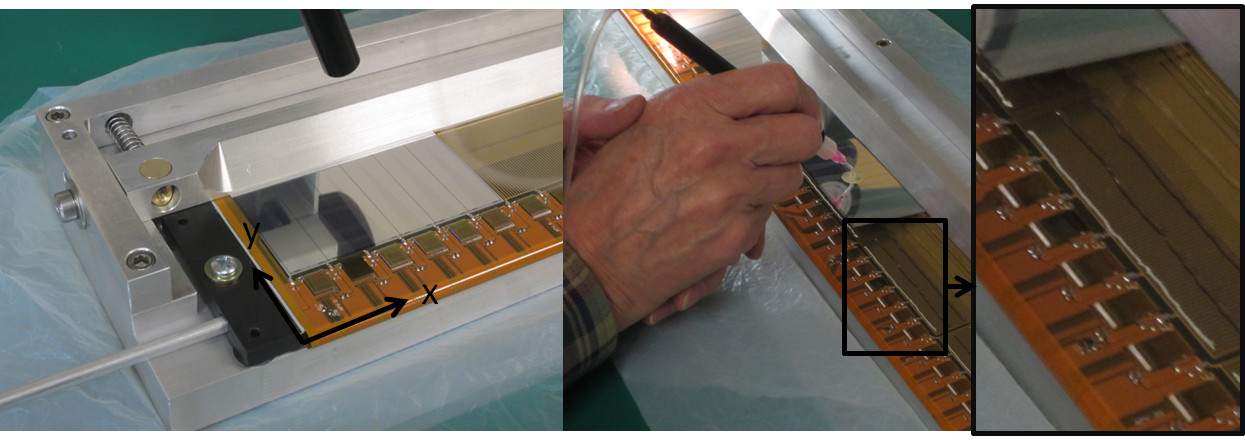}
\caption{Left shows the silicon sensor attach tooling.  The right two photos show
a sensor being placed and highlight the two different types of epoxy used.
\label{fig:si_glue}}
\end{center}
\end{figure}

The sensors are held down with two different types of epoxy as shown in figure~\ref{fig:si_glue}. The first is a
conductive epoxy needed to make a good contact between the aluminized back side
of the sensor and the traces below, which carry the bias voltage. We used the
same TRA-DUCT 2902 adhesive for this. The second epoxy is EPON 828 resin with
VERSAMID 140 hardener. We used this epoxy for mechanical strength and stability since the
conductive epoxy is not very strong. A single line of conductive epoxy was put
down and then several lines of conventional epoxy were applied to make a strong
bond. We found that the working time after applying the glue was about 30
minutes after which it would be much more difficult to get the sensor to lie
flat.

\begin{figure}[h]
\begin{center}
\includegraphics[width=2in, keepaspectratio=true, angle=0]{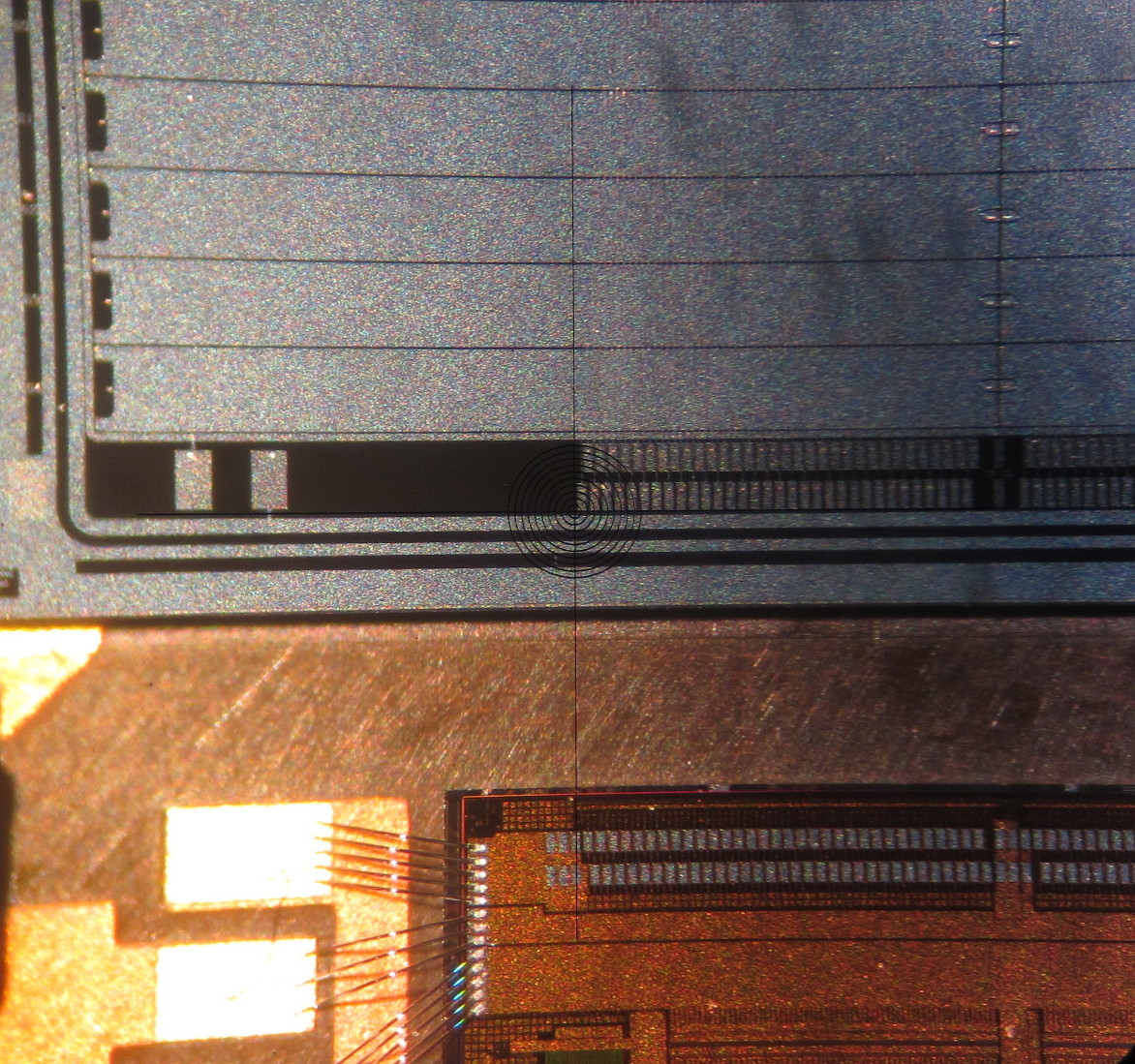}
\caption{The reticule in the microscope is used to align the bonding pads
between a sensor and the APV chips which serve it.
\label{fig:si_align}}
\end{center}
\end{figure}

The silicon sensors were aligned in a similar way to the APV chips. The tooling
provided a stop in the ``y'' direction so the technician needed only focus on
alignment in the ``x'' direction. Under a microscope, a technician would align the
sensor bonding pads with the APV bonding pads checking all 6 APV chips for each
sensor. An example view through the microscope during this alignment procedure
is shown in figure~\ref{fig:si_align}. After all sensors were aligned, weights were applied to the sensors.
In early prototypes we found that the sensors would drift during the curing
process. Adding weights to the tops of the sensors prevented the sensors from
drifting. After the weights were applied, the alignment was rechecked and
adjusted if necessary. Staves would be allowed to cure for 24 hours before
they would be packaged and transported.

\section{Silicon sensor bonding and testing}
\begin{figure}[h]
\begin{center}
\includegraphics[width=6.5in, keepaspectratio=true, angle=0]{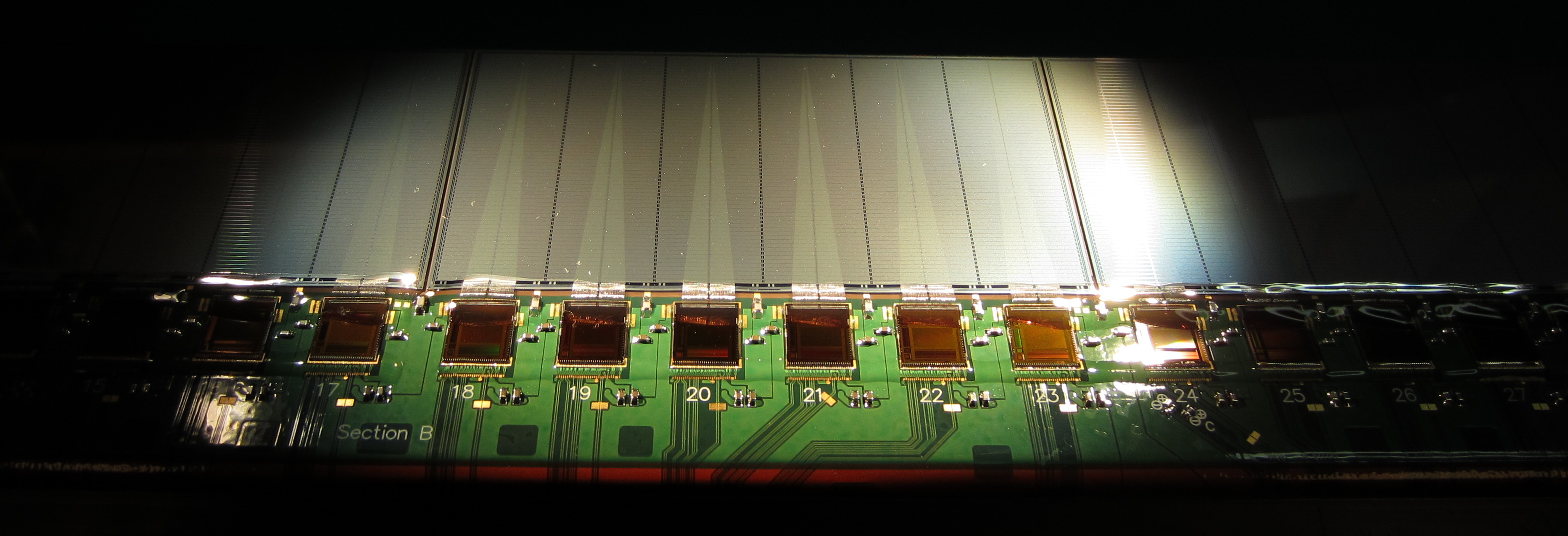}
\caption{Closeup of a sensor on a completed stave.  The encapsulant is difficult
to see because of its transparency.
\label{fig:si_bonded}}
\end{center}
\end{figure}

\begin{figure}[h]
\begin{center}
\includegraphics[width=6.5in, keepaspectratio=true, angle=0]{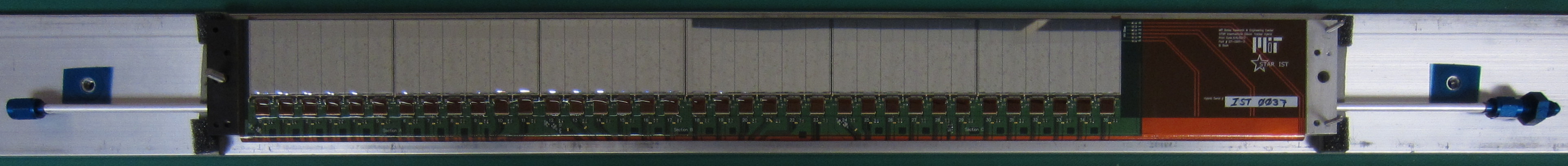}
\caption{A completed stave ready for installation.
\label{fig:full_stave}}
\end{center}
\end{figure}

The bonding of the silicon sensors took place at the same Instrumentation
Division at BNL on the same bonding equipment. Again \SI{1}{\mil} aluminum wire was
used. An automatic bonding program was used but we required close supervision by
the technician. Because of the size and pitch of the bonding pads it was
important to catch any error and repair it before bonding continued. The
bonding process put down 128 bonds between each APV chip and each silicon sensor
for a total of 4608 wire bonds. In addition, 24 bonds were put down between
each sensor and the hybrid as part of the biasing circuitry for a total of 144
bonds. Depending on the amount of technician intervention, this process would
take between 1 and 2 hours.  The end result can be seen in figure~\ref{fig:si_bonded}
and figure~\ref{fig:full_stave}.

Fully bonded staves were brought back into the IST clean room and tested
again. Using the DAQ system, we were able to identify any shorted bonding wires
and determine if it was possible to repair them. In addition, baseline noise
and pedestal information was again recorded. We found very few errors with the
bonding done at BNL and nearly all of the staves produced are in use in the
final detector. Fully functioning staves then had an encapsulant applied to the
bonding wires to protect them during handling and installation. The first
prototypes used Dymax 9001-E-V3.1 cured with a Dymax 5000-EC series ultraviolet
curing lamp. We found that, after exposure to the UV light source, some APV chips
would no longer function or have higher input noise. While the exact cause of
this was unknown, we suspect that the silicon was damaged by the UV treatment.
For production staves, Dow Corning Sylgard 186 elastomer was used. This
encapsulant has a room temperature cure and did not affect the performance of
the APV chips.  A fully completed stave, ready for installation, is shown in
figure~\ref{fig:full_stave}.

After the encapsulation the staves were tested extensively and optically surveyed
before being installed.

\section{Optical survey}
\begin{figure}[h]
\begin{center}
\includegraphics[height=2in, keepaspectratio=true, angle=0]{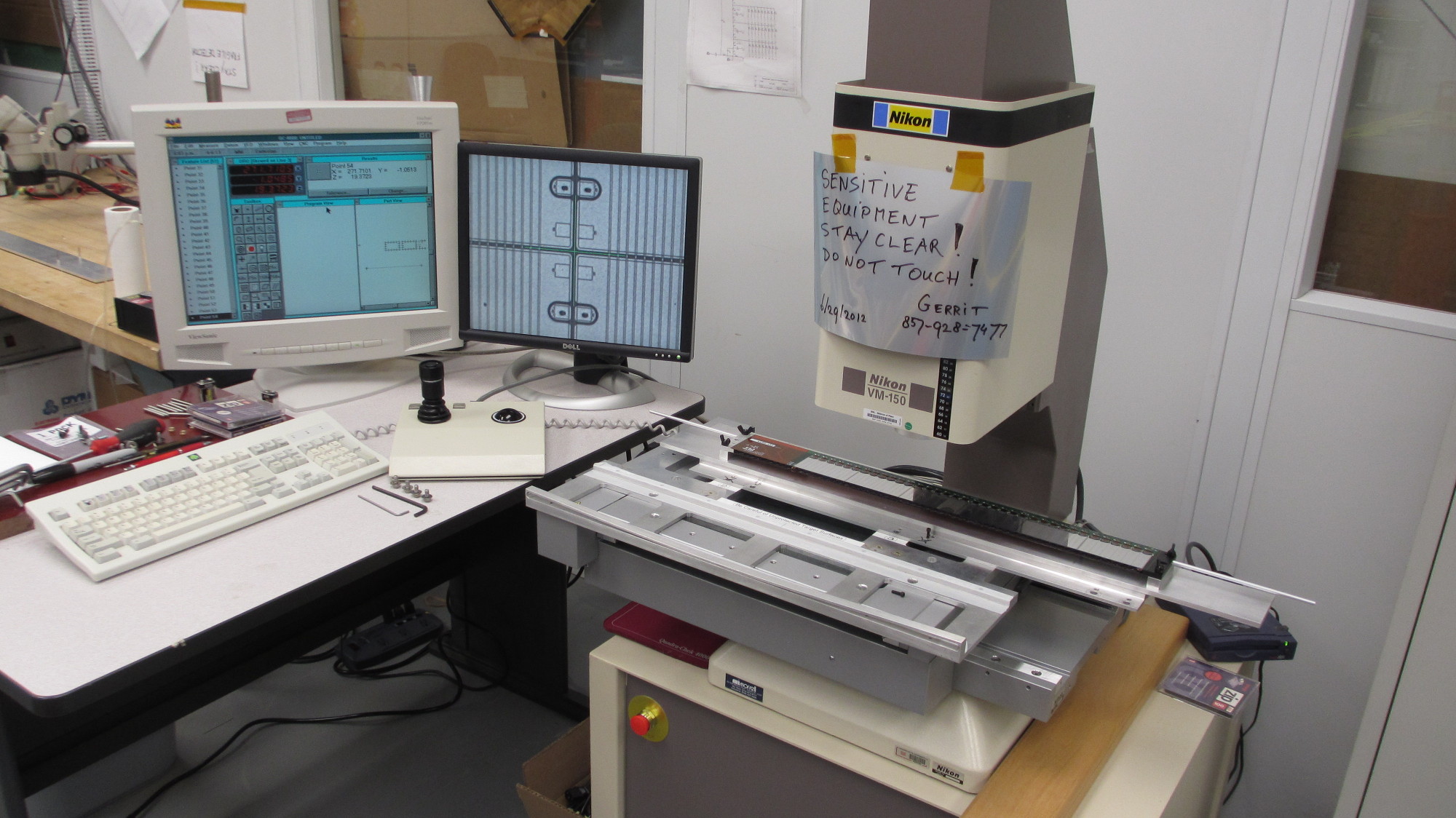}
\includegraphics[height=2in, keepaspectratio=true, angle=0]{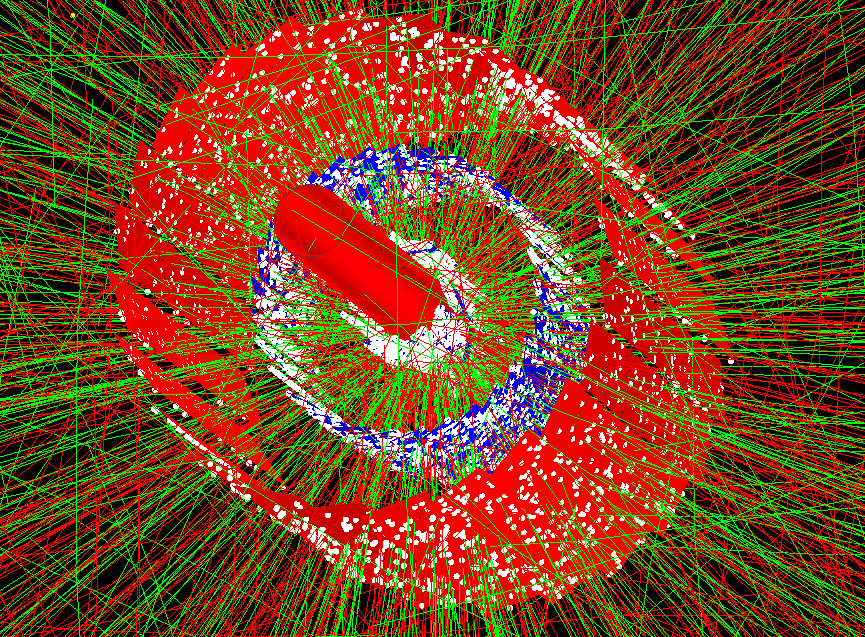}
\caption{Left shows a stave on the survey station.  Right shows a computer
reconstruction of tracks going through the IST (red layer) and other detectors which
surround it.
\label{fig:survey}}
\end{center}
\end{figure}

The structure which the staves are mounted to has a mechanical alignment
accuracy of about \SI{200}{\micro\meter}. However, once installed we require knowing the
position of the detector to a higher degree of accuracy. We are able to do this
by aligning tracks which pass through the IST as well as other detectors. We
can reduce the number of tracks needed to do this alignment by knowing very
precisely how the sensors on a single stave are located relative to each other.
Before installation each of the staves was placed on an optical survey station
and measured. The survey station, a Nikon VM-150, was used to measure each of
the staves. We estimate the accuracy of these measurements to be about \SI{5}{\micro\meter} in
the plane of the detector and \SI{50}{\micro\meter} out of plane. We combine this information
along with the ``rough'' position of each detector inside the experiment and will
be able to align the detector from tracks generated in this and other detectors
(see figure~\ref{fig:survey}).

\section{Installation}
\begin{figure}[h]
\begin{center}
\includegraphics[width=5in, keepaspectratio=true, angle=0]{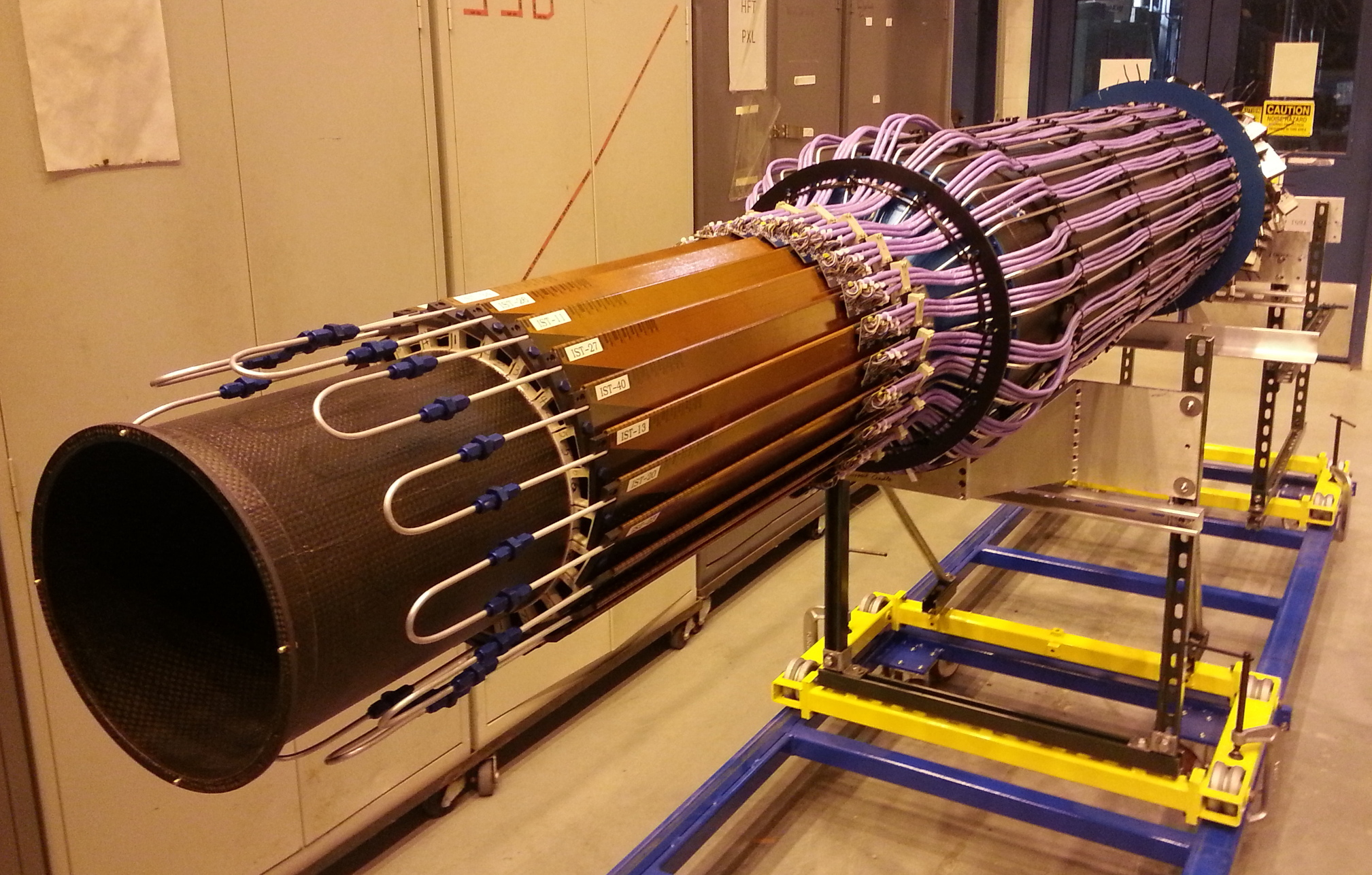}
\caption{Completed IST detector ready for insertion into STAR.
\label{fig:full}}
\end{center}
\end{figure}

Fully functional staves were installed on a carbon fiber support structure
arranged to have full azimuthal coverage. As seen in figure~\ref{fig:full} The silicon components are facing
into the support structure which protects the silicon from damage during the
full installation process. Electrical connections were laid on the tube and
connected to each of the detectors. We were able to run the detectors in place
to ensure that all electrical connections were sound. Data was taken to establish
the baseline noise and pedestal data before final installation. Finally, cooling
lines were attached to the cooling tubes running through the staves and were
tested with a helium leak checker. Two staves were found to have leaks inside
the cooling tube. These staves were not installed and the cause of the leaks is
currently being investigated.

Once the entire IST was assembled the support structure was mated with a
larger detector support system and inserted into the central region of STAR.
The detector was then hooked up to electrical and cooling services and run in
place again to ensure that all detectors are working properly.

\section{Summary}
The IST for STAR is made of 24 staves, each of which carry 6 silicon sensors read
out by 36 APV chips. The detectors are designed to operate in a high radiation
environment with as little mass as possible to reduce the chance of interactions
with particles. The staves were designed and manufactured at MIT and LBNL.
Passive assembly as well as APV readout chip and silicon sensor attachment took
place at MIT while bonding, testing and installation took place at Brookhaven
National Laboratory. The finished detector was installed into STAR and is
currently taking data with 95 percent of channels working properly.

\section{Acknowledgements}
The fabrication of the IST was split between MIT and University of Illinois at Chicago.
Dr. Zhenyu Ye at UIC coordinated with the Silicon Detector Lab at Fermi National
Accelerator Laboratory to complete all steps in production after the passive
attach.  Their assembly process was very similar to the assembly process used
at MIT but not identical; it is outside the scope of this paper to detail the
two separate assembly processes.

\end{document}